\documentstyle[11pt,epsf]{article}
\newlength{\bredde}
\def\slash#1{\settowidth{\bredde}{$#1$}\ifmmode\,\raisebox{.15ex}{/}
\hspace*{-\bredde} #1\else$\,\raisebox{.15ex}{/}\hspace*{-\bredde} #1$\fi}
\newcommand{\beq}{\begin{equation}}
\newcommand{\eeq}{\end{equation}}
\newcommand{\noi}{\vspace{12pt}\noindent}
\newcommand{\lG}{\raise.3ex\hbox{$\stackrel{\leftarrow}{G}$}}
\newcommand{\lU}{\raise.3ex\hbox{$\stackrel{\leftarrow}{U}$}}
\newcommand{\lP}{\raise.3ex\hbox{$\stackrel{\leftarrow}{{\cal P}}$}}
\newcommand{\leta}{\raise.3ex\hbox{$\stackrel{\leftarrow}{\eta}$}}
\newcommand{\lOmega}{\raise.3ex\hbox{$\stackrel{\leftarrow}{\Omega}$}}
\newcommand{\ldr}{\raise.3ex\hbox{$\stackrel{\leftarrow}{\delta^r}$}}
\def\beqn{\begin{eqnarray}}
\def\eeqn{\end{eqnarray}}

\def\gtwid{\raise.3ex\hbox{$>$\kern-.75em\lower1ex\hbox{$\sim$}}}
\def\ltwid{\raise.3ex\hbox{$<$\kern-.75em\lower1ex\hbox{$\sim$}}}

\begin{document}
\topmargin -1.4cm
\oddsidemargin -0.8cm
\evensidemargin -0.8cm
\title{\Large{A Quark-Antiquark Condensate in Three-Dimensional QCD}}

\vspace{0.9cm}

\author{
{\sc P.H. Damgaard$^a$, U.M. Heller$^b$, A. Krasnitz$^c$ and T. Madsen$^a$}
\\~\\~$^{a)}$ The Niels Bohr Institute\\Blegdamsvej 17\\DK-2100
Copenhagen\\Denmark\\~\\~$^{b)}$ SCRI\\Florida State University \\
Tallahassee, FL 32306-4130\\USA\\~\\~$^{c)}$Unidade de Ci\^encias
Exactas e Humanas\\Universidade do
Algarve \\ Campus de Gambelas, P-8000 Faro \\ Portugal
}
 
\maketitle
\vfill
\begin{abstract} 
Three-dimensional lattice QCD is studied by Monte Carlo simulations
within the quenched approximation. At zero temperature a quark-antiquark
condensate is observed in the limit of vanishing quark masses. The
condensate vanishes continuously at the finite-temperature deconfinement 
phase transition of the theory. A natural interpretation of this phenomenon
in the full theory with dynamical quarks is in terms of the spontaneous
flavor symmetry breaking $U(N_f) \to  U(N_f/2)\times U(N_f/2)$.
In addition, the spectrum of low-lying Dirac operator eigenvalues is
computed and found to be consistent with a flat distribution at zero
temperature, in agreement with analytical predictions.
\end{abstract}
\vspace{0.3cm}

\vfill
\begin{flushleft}
NBI-HE-98-06 \\
FSU--SCRI--98--28\\
UALG/TP/98-4 \\
hep-lat/9803012
\end{flushleft}
\thispagestyle{empty}
\newpage

\noindent
Although super-renormalizable, non-Abelian gauge theories in (2+1) dimensions
are notoriously difficult to analyze in perturbation theory due
to the strong infrared divergences that appear already at leading orders.
For this reason, a very appropriate framework for the study of such
three-dimensional theories is that of lattice gauge theory \cite{DKK}. 
In this letter we shall report
on a series of Monte Carlo simulations aimed at clarifying the issue of
flavor symmetry breaking and, at high temperature, flavor symmetry
restoration in three-dimensional QCD.

\noi
One striking difference between massless (2+1)-dimensional QCD 
and its (3+1)-dimensional counterpart is the absence of 
chiral symmetry. The lowest representation of the Clifford
algebra in three dimensions has dimension two. Spinors are thus naturally
taken to be two-spinors, and as $\gamma$-matrices one can choose the three 
Pauli matrices $\sigma_i$. On the surface there seems to be no room for 
an additional analogue of
$\gamma_5$, and hence no chiral symmetry. A fermionic mass term does,
however, break parity ${\cal P}$ and time-reversal ${\cal T}$ symmetries
\cite{JT}, symmetries that are both respected by the massless Lagrangian.
By introducing an {\em even} number of quark species $N_f$, and combining
the spinors in pairs into four-spinors $\psi$, one can rewrite the
fermionic part of the Lagrangian entirely in terms of three 4-dimensional 
$\gamma$-matrices $\gamma_0, \gamma_1, \gamma_2$ and the four-spinors $\psi$.
A mass term of the kind $m\bar{\psi}\psi$ can now be introduced \cite{JT,P}.
Such a mass term is ${\cal P}$ and ${\cal T}$
invariant. It does, however, break both of the ``chiral'' symmetries
associated with $\gamma_4$ and $\gamma_5$ rotations. If we restrict our
attention to an even number of massless quark flavors we thus have a 
formulation which is very similar to that of four-dimensional massless
QCD. There are chiral symmetries, and the strong, in the absence of matter 
fields almost certainly confining, gauge interactions. For a small even number 
of flavors one could expect these chiral symmetries to break spontaneously.

\noi
It has been suggested that a plausible flavor symmetry breaking in the
theory with an even number of flavors $N_f$ should be that of 
$U(N_f) \to  U(N_f/2)\times U(N_f/2)$ \cite{P,VZ}. The formation of
a chiral condensate $\langle\bar{\psi}\psi\rangle$ would be a signal for
such spontaneous symmetry breaking. If correct, it would open up the
possibility of comparing the detailed universality predictions \cite{VZ,ADMN} 
for the Dirac operator spectrum around eigenvalues near $\lambda\!=\! 0$
with lattice Monte Carlo data, as has recently been done in the case
of four-dimensional lattice QCD \cite{BMSVW}. The first step in such a
program is to establish the existence of the chiral condensate
$\langle\bar{\psi}\psi\rangle$.

\noi
The transcription of the formulation in terms of four-spinors and
four-dimensional $\gamma$-matrices to a lattice theory is quite
naturally done in terms of staggered fermions \cite{BB}. Interpreting
three-dimensional staggered fermions in this manner has already been
explored in several Monte Carlo analyses, see $e.g.$ ref. \cite{HKK}.
The standard three-dimensional action for free staggered fermions,
keeping the lattice spacing $a$ explicitly,
\beq
S_{F}+S_{M} = - \frac{1}{2} a^{d-1} \sum_{r,\mu} (-1)^{r_1+r_2+...+r_{\mu-1}}
[\overline{\chi}(r)\chi(r+\mu)+ \overline{\chi}(r+\mu)\chi(r)]+ima^d \sum_r 
\overline{\chi}(r)\chi(r) 
\eeq
can be transformed into a form with 2-component Dirac fields. In the limit 
$a{\rightarrow}0$ it will go to the continuum action for free Dirac fermions 
with two flavors and (if the kinetic terms is defined with respect to the
same $\gamma$ matrices; see below) masses $m$ and $-m$. In three 
dimensions there are two equivalence classes of irreducible 
representations of the Clifford algebra, 
${\{\gamma_\mu\}}$ and ${\{-\gamma_\mu\}}\equiv{\{\beta_\mu\}}$.

\noi
Following ref. \cite{BB} the new action can be defined on a  
lattice with twice the original lattice spacing:
\begin{eqnarray*}  
S_F+S_M  =  (2a^d)\sum_{r,\mu} \bigg\{\overline{u}(\gamma_\mu \otimes I)
\partial_\mu u+\overline{d}(\beta_\mu \otimes I)\partial_\mu d+
a[\overline{u}(I \otimes \gamma^{T}_{\mu})\partial_{\mu}^{2}d+
\overline{d}(I \otimes \beta^{T}_{\mu})\partial_{\mu}^{2}u]   
\end{eqnarray*}
\beq
+ im[\overline{u}(I \otimes I)u+\overline{d}(I \otimes I)d]\bigg\}\label{t1}  
\eeq 

\noi
Here $u$ and $d$ are 2-component spinors and in the quark bilinears the first 
$2\times2$ matrix acts on spinor indices while the second acts on flavor 
indices. One can identify a $U(1)\times U(1)$ symmetry in eq. (\ref{t1}) as
\beq
{u \choose d}\rightarrow e^{i\theta_1}{u \choose d}, \quad(\overline{u},
\overline{d})\rightarrow({\overline{u}, \overline{d}})e^{-i\theta_1}\label{t2}
\eeq
\beq
{u \choose d} \rightarrow {{\cos{\theta_2} \quad i\sin{\theta_2}} \choose 
{i\sin{\theta_2} \quad \cos{\theta_2}}}{u \choose d},\quad (\overline{u},
\overline{d}) \rightarrow({\overline{u}, \overline{d}}) {{\cos{\theta_2} 
\quad i\sin{\theta_2
}} \choose {i\sin{\theta_2} \quad \cos{\theta_2}}}.\label{t3}
\eeq
The symmetry in eq. (\ref{t3}) is a remnant of the three-dimensional
``chiral symmetry'' on the lattice, and 
it is broken by the inclusion of a mass term. The symmetry of eq. (\ref{t2}) 
is nothing but fermion number conservation.

\noi
We consider gauge group $SU(3)$, and add to the fermionic part of the
action the conventional Wilson gauge action ($\beta \equiv 6/g^2$):
\beq
S_{\mbox{\rm gauge}} = \beta \sum_p (1 - \frac{1}{3}{\mbox{\rm Re Tr}}U_p) ~,
\eeq
where $U_p$ is the product of links along a fundamental plaquette. 
The canonical ensemble of gauge field configurations was generated using a
combination of microcanonical over-relaxation (MOR) \cite{MOR} and
quasi-heat-bath (QHB) algorithms \cite{K_P}. An individual MOR or QHB step
consisted of updating 
consecutively the three $2\times 2$ submatrices of the SU(3) link variable
\cite{cabmar}. In our implementation five MOR steps were followed by
two QHB ones.

\noi
We now present the results of our Monte Carlo simulations. Instead of first
considering symmetric (2+1)-dimensional lattice volumes, we have immediately
turned to lattices with the temporal direction much smaller than the two
spatial directions. This has allowed us to study the finite-temperature 
behavior of the theory, while at the same time determining, in the 
low-temperature region, whether at all a quark-antiquark condensate is 
formed at zero temperature. As mentioned in the introduction, all
simulations have been performed in the quenched approximation. This should
have no bearing on the qualitative features as compared to the full theory
with dynamical quarks, at least as long as the number of flavors is
relatively small. We have monitored all relevant gauge-invariant correlation
functions in both the matter and gauge sector, and in particular of course
the crucial order parameters: the Polyakov line $\langle W\rangle$ and the 
chiral condensate $\langle\bar{\psi}\psi\rangle$. Since the gauge dynamics
is unaffected by the fermions in this approximation, the theory should
undergo a 2nd-order finite-temperature deconfining phase transition at
an intermediate $\beta$-value \cite{CTDW}. This particular detail may of
course be modified by the presence of (light) dynamical quarks (just as
its four-dimensional counterpart), but the qualitative feature of a switch
from confining to deconfined characteristics should persist.

\begin{figure}
\epsfxsize=5.0in
\centerline{\epsfbox[0 0 576 376]{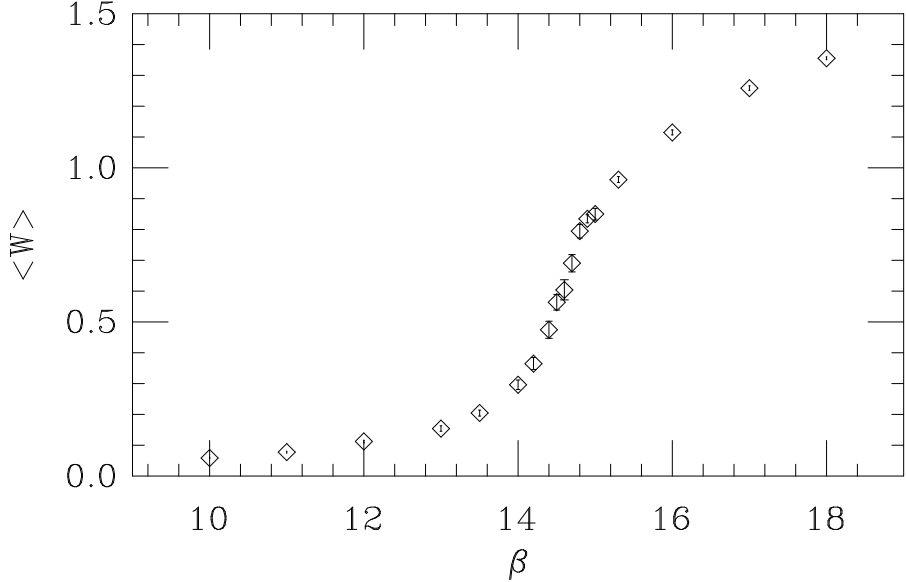}}
\caption{The $\beta$ dependence of the Polyakov line $\langle W\rangle$,
measured on a $40^2\times 4$ lattice.}
\label{wplot}
\end{figure}

\noi
The finite-temperature phase transition in the gauge sector is associated
to global $Z(3)$ symmetry breaking. The Polyakov line aligns, in the
deconfined phase, along the three complex roots of $z^3\!=\!1$ on the
unit circle, $i.e$ along $z \in \{1,\exp[2i\pi/3],\exp[4i\pi/3]\}$. In
the theory with dynamical quarks, these act like a magnetic field that
force the spontaneous symmetry breaking on the Polyakov line to occur
on the physical real axis ($i.e.$ along $z\!=\!1$) \cite{BHDG}, but in this
quenched situation there is equal probability for the Polyakov line to fall
into any of the three sectors. This is normally not a question of great
concern, since in finite-volume simulations one in any case needs a
modified working definition of the Polyakov line expectation value,
which customarily is based on taking its absolute value. But as has recently
been emphasized \cite{CC}, the existence of
complex phases in the Polyakov line condensate can have strong effects
on the chiral condensate in quenched simulations.\footnote{The analogous
phenomenon for $SU(2)$ (where it just amounts to a sign, and in fact can
effectively flip the fermion boundary conditions from antiperiodic to
periodic), was already pointed out in ref. \cite{Ketal}.} We have seen this
very clearly in our simulations. In Figure \ref{wplot} we display the (absolute
value) of the Polyakov loop expectation value on a $40^2\!\times\! 4$ lattice,
as a function of the gauge coupling $\beta$. It shows the expected behavior
of a smooth phase transition at an intermediate value of $\beta$, which we 
read off to be around $\beta\!\sim\! 15$. This is in good agreement with
earlier Monte Carlo results \cite{Lege} and mean field theory \cite{Billo}.
It is also in rough agreement with
three-dimensional scaling, and earlier Monte Carlo results for a temporal
extent of $N_{\tau}\!=\! 2$ which gave a critical $\beta$-value of
$\beta_c \simeq 8.1$ \cite{CTDW}. If we re-instate the lattice spacing $a$,
the definition of $\beta$ actually becomes $\beta\!\equiv\! 6/(g^2a)$,
that is, the coupling $g$ is dimensionful in three dimensions. With
temperature given by $T\!=\!1/(N_{\tau}a)$, perfect scaling in this theory
would imply that for the physical critical temperature $T_c$ to stay
fixed, the critical coupling $\beta_c$ should be $2\times8.1\!=\!16.2$
on our lattice with $N_{\tau}\!=\!4$. Thus, although continuum scaling is not 
perfectly obeyed here, the deviation is nevertheless only around 5\%.

\begin{figure}
\epsfxsize=5.0in
\centerline{\epsfbox[0 0 576 376]{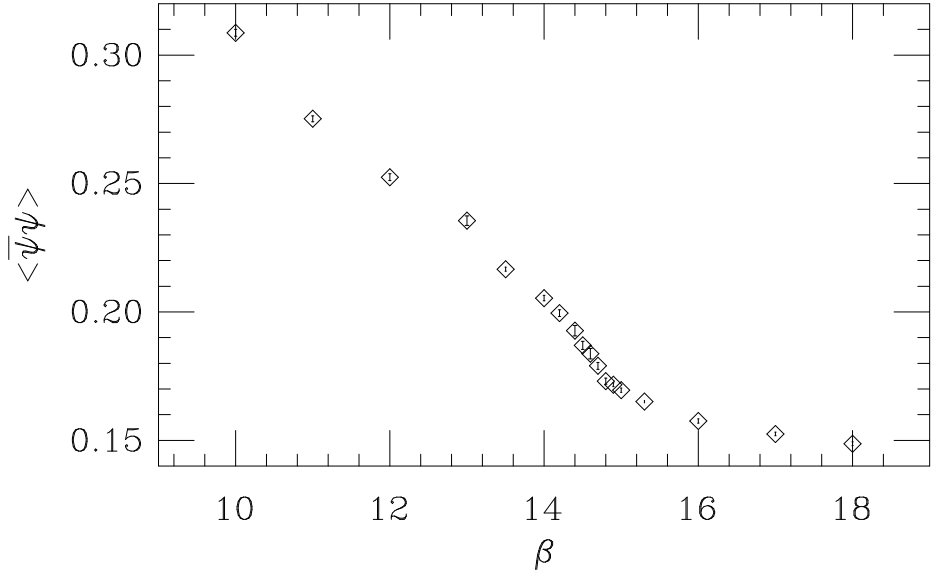}}
\caption{The $\beta$ dependence of the quark-antiquark condensate
$\langle{\overline\psi}\psi\rangle$, for $am_q=0.1$,
measured on a $40^2\times 4$ lattice.}
\label{pbplot}
\end{figure}

\noi
While the behavior of the Polyakov line thus is completely as expected,
the condensate $\langle\bar{\psi}\psi\rangle$ displayed at first a quite 
surprising behavior, which we eventually could attribute to the complex
phases of the Polyakov line. Before presenting details on this particular
issue, we first display, in Figure \ref{pbplot}, a typical plot of our final 
results (here for a quark mass of $am_q=0.1$). This figure, reminiscent of
similar plots for SU(2) in 4 dimensions \cite{Ketal}, 
suggests immediately that,
in the limit of massless quarks, (1) a condensate is formed in the
zero-temperature theory (corresponding to the phase with $\beta<\beta_c$),
but not at high temperature -- we shall confirm these observations with
explicit extrapolations to the massless limit below -- and (2) at the
deconfinement transition this condensate approaches zero smoothly. Neither of
these observations come as a surprise. First of all, the formation of a
$\bar{\psi}\psi$-condensate
at strong coupling is almost built in by the present lattice formulation.
The strong-coupling, zero-temperature analysis of ref. \cite{KS} 
straightforwardly carries over to the present (2+1)-dimensional case, 
where it again predicts
the formation of a chiral condensate.\footnote{The prediction is not as strong
as in (3+1) dimensions, however, since one must view it in terms of a
$1/d$-expansion that quantitatively could fail for $d\!=\!3$.} Variational
calculations in the Hamiltonian formalism \cite{Luo} indicate the existence
of the condensate also at finite couplings.
Similarly,
the extension to finite temperature \cite{DKS} immediately predicts, at
strong coupling, a symmetry-restoring phase transition at finite temperature.

\begin{figure}
\epsfxsize=6.0in
\centerline{\epsfbox[0 0 576 576]{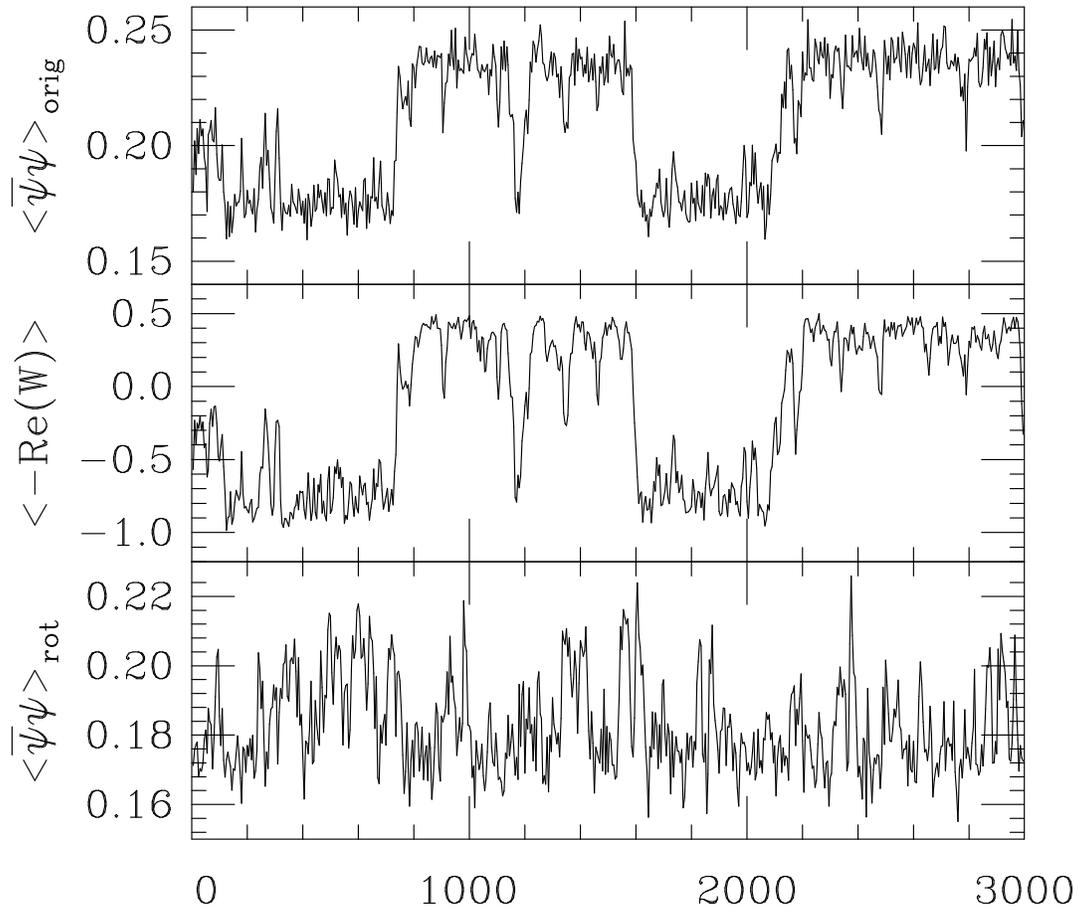}}
\caption{The Monte Carlo time history, averaged over 5 consecutive iterations,
of the quark-antiquark condensate
$\langle{\overline\psi}\psi\rangle$ for $am_q=0.1$ measured on the original
configurations (top), of the negative of the real part of $\langle W\rangle$
(middle) and of $\langle{\overline\psi}\psi\rangle$ measured on the
transformed configurations (bottom)
at $\beta=14.6$ on a $40^2\times 4$ lattice.}
\label{opbph}
\end{figure}

\noi
Of course, details will depend on the quenched approximation. For instance,
it is almost impossible to imagine that in the presence of a continuous
deconfining phase transition the behavior of $\langle\bar{\psi}\psi\rangle$
could be anything but smooth in our case. To show that this is not what
is found if one blindly measures $\langle\bar{\psi}\psi\rangle$ in this
quenched simulation, we show in Figure \ref{opbph} (top) a typical run history 
of that quantity. For illustrative purposes we have here chosen 
$\beta\!=\!14.6$ associated with $Z(3)$ breaking. Surprisingly, the chiral 
condensate displays what appears to be a clear-cut two-state signal,
indicating the proximity of a {\em discontinuous} phase transition.
The reason for this bizarre behavior becomes evident if we compare (see
Figure \ref{opbph} (middle)) with
the corresponding time history of the (negative of the) real part of 
the Polyakov line,
$\langle{\mbox{\rm - Re}}W\rangle$. The apparent ``two-state signal'' 
appears to be
reproduced in this variable, and in fact the data in the upper and middle
part of Figure \ref{opbph} are
almost perfectly correlated. We recall that in the case of the
Polyakov line, this behavior simply reflects the flipping around between
the three different $Z(3)$ phases, and in no way implies a two-state
signal in the properly defined Polyakov line. Furthermore,
the anti-correlated behavior
between $\langle{\mbox{\rm Re}}W\rangle$ and  $\langle\bar{\psi}\psi\rangle$
is in complete agreement with the observations of ref. \cite{CC}:
When the Polyakov line lies along one of its two complex phases (and its real
part hence is relatively small), the chiral condensate has an unphysical
{\em larger} value than in the physical situation corresponding to
a Polyakov line aligned along the real axis.

\noi
The problem described above can be solved in different ways. We have opted
for the following simple method. The phase $\phi$ of $\langle W\rangle$
is measured. Then, if $\phi >\pi/3$, all the timelike link variables
in the $t=0$ time slice are multiplied by $\exp(-2\pi i/3)$. If $\phi <-\pi/3$,
these variables are multiplied by $\exp(2\pi i/3)$. If we find 
$-\pi/3\leq\phi\leq\pi/3$, the configuration is left unchanged. 
Thus, for the transformed configuration, the phase of $\langle W\rangle$ lies
between $-\pi/3$ and $\pi/3$. The 
transformation as described is an exact symmetry of the quenched theory.
Nevertheless, we prefer to discard the transformed configuration after
performing measurements, and to continue the Monte Carlo process with the
original configuration.
In the bottom part of Figure \ref{opbph}
we show how the time history of $\langle\bar{\psi}\psi\rangle$ looks with
this prescription. We have here used precisely the same starting 
configuration, and yet the result is radically 
different. There is now no trace of any metastability, and
$\bar{\psi}\psi$ is nicely fluctuating around one particular, stable, value.
We have applied this prescription wherever we detected a two-state signal in the
$\langle \bar{\psi}\psi\rangle$ time history. This resulted in a smooth 
dependence of $\langle \bar{\psi}\psi\rangle$ on $\beta$, as shown in Figure 
\ref{pbplot}.\footnote{The used prescription is, however, not entirely free of
systematic error: 
to every transition between the physical and the unphysical values
of $\bar{\psi}\psi$ in Figure \ref{opbph} corresponds a spike in the refined
value of $\bar{\psi}\psi$. As a result of these spikes,
we expect the average $\langle \bar{\psi}\psi\rangle$ to be slightly
biased upwards. This bias can be systematically reduced by increasing 
the lattice volume,
thereby making the transitions between the phases of $\langle W\rangle$ 
increasingly rare.}

\begin{figure}
\epsfxsize=4.0in
\centerline{\epsfbox[0 0 576 576]{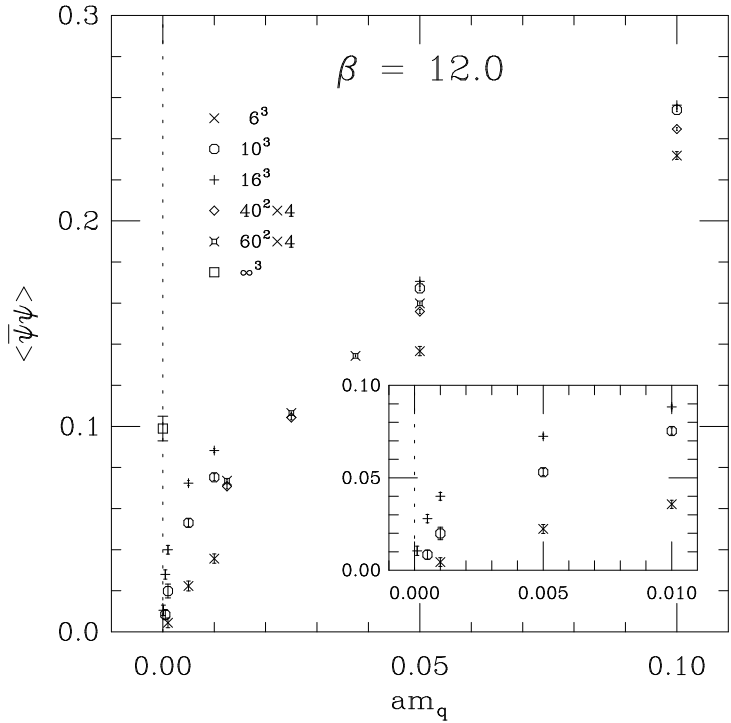}}
\caption{The quark-antiquark condensate $\langle\bar{\psi}\psi\rangle$ as a 
function of the quark mass $am_q$ at $\beta=12.0$, corresponding to 
the low-temperature phase, for various volumes. Also shown is the value
for the condensate at $am_q=0$ in the infinite volume limit (square).}
\label{pbpvsm}
\end{figure}

\begin{figure}
\epsfxsize=4.0in
\centerline{\epsfbox[0 0 576 576]{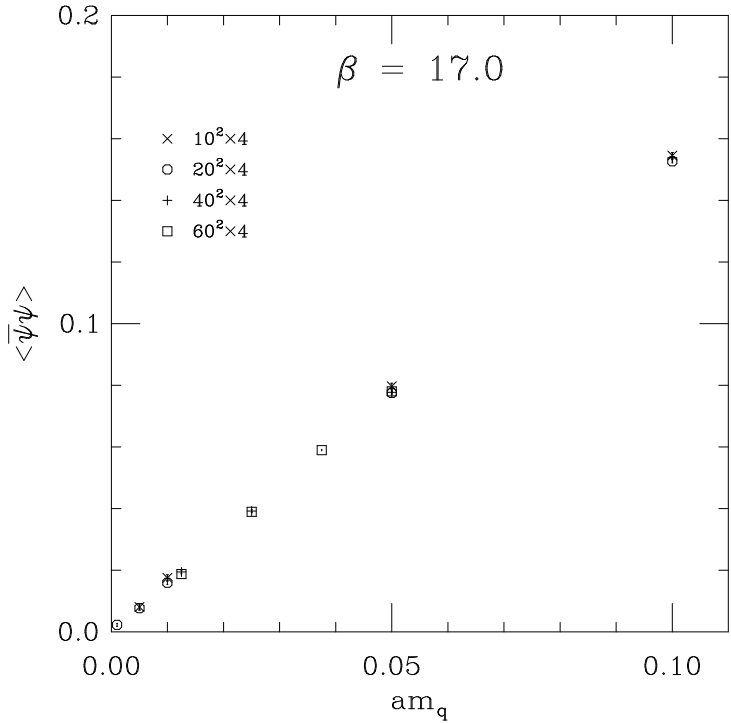}}
\caption{Same as Figure \protect\ref{pbpvsm} but in the high-temperature
phase, at $\beta=17.0$.} 
\label{pbpvsm1}
\end{figure}

\noi 
While the data already shown indicate the presence of a
non-vanishing quark-antiquark condensate at low temperature, we have 
performed a much more detailed analysis to really confirm this result.
The problem of course is that for any finite volume $V$ there is no
spontaneous symmetry breaking in the $am_q \to 0$ limit. Fortunately
the behavior with increasing volume is, however, quite different in
a phase of broken symmetry compared with a phase of restored symmetry.
We can use this different behavior to test the hypothesis of spontaneous
symmetry breaking in the confined phase of the theory. In Figure \ref{pbpvsm}
we display data $\langle\bar{\psi}\psi\rangle$ for a variety of lattice 
volumes at $\beta = 12.0$ (corresponding to the confined phase). While
the condensate eventually goes towards zero as $am_q\to 0$ for any of
the lattices volumes, there is a clear trend towards increasing values,
at fixed $am_q$, as the volume is increased. On the same figure we have
also indicated the value of the condensate
at $am_q=0$ and extrapolated to $V=\infty$
as obtained from an entirely different analysis (see below). Considering the
errors, this extrapolated value is completely consistent with the direct
measurements at increasingly larger volumes, as shown in Figure \ref{pbpvsm}.
It finally remains to be tested if the behavior on the high-temperature 
side is consistent with a vanishing condensate in the limit of zero quark 
masses. In Figure \ref{pbpvsm1}
we show the convergence of $\langle\bar{\psi}\psi\rangle$ at $\beta=17.0$. 
These data are clearly very different from those in the confined phase,
with no discernible volume dependence. All points lie on a straight line
that directly extrapolates to a vanishing condensate at $am_q=0$.

\begin{figure}
\epsfxsize=6.0in
\centerline{\epsfbox[0 0 576 576]{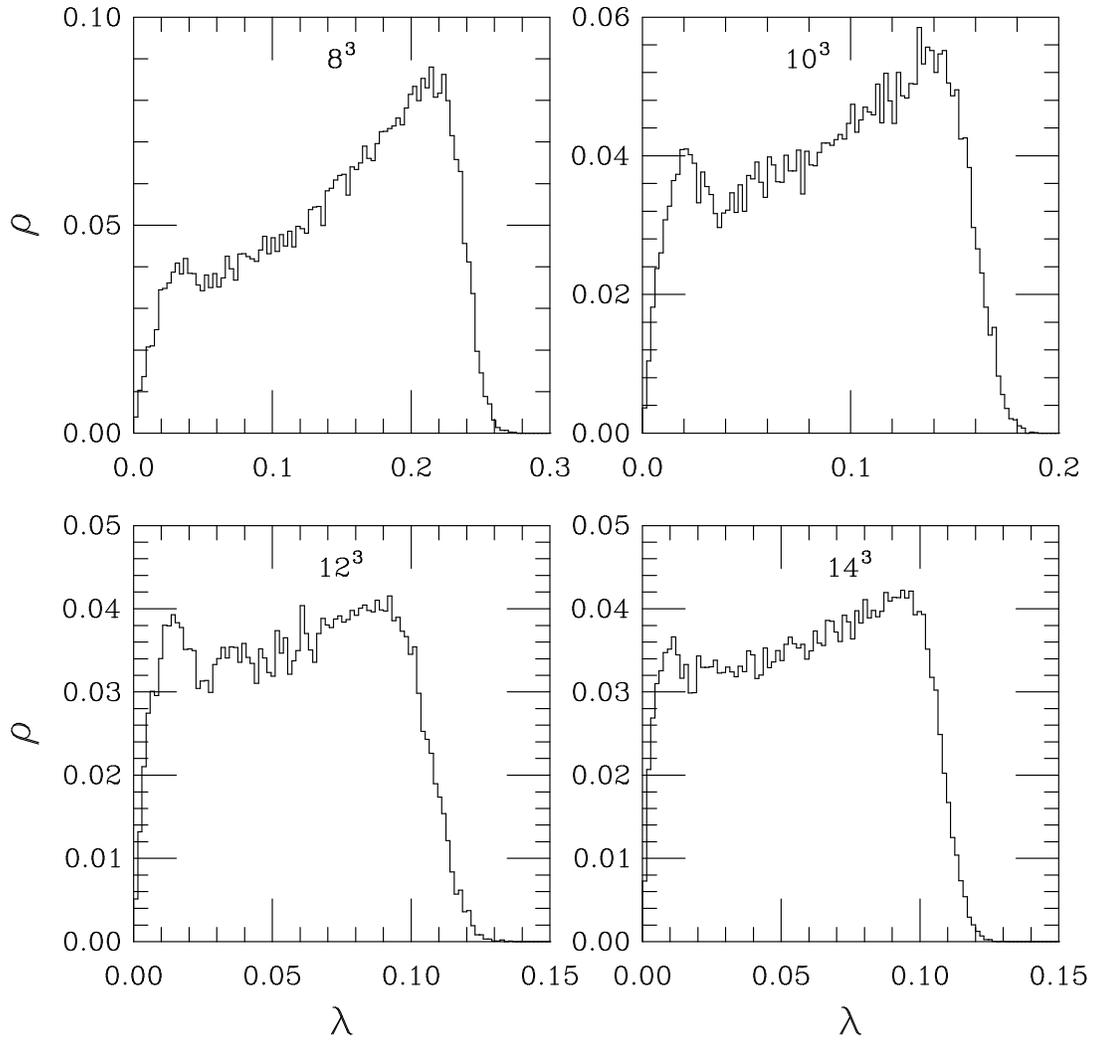}}
\caption{The density of low lying eigenvalues for various volumes
at $\beta=12.0$.} 
\label{histogr}
\end{figure}

\noi
We have shown that, at least in the quenched approximation and in
our present latticized version, a chiral condensate {\em is} formed
in (2+1)-dimensional QCD. This condensate disappears
according to expectations at the finite temperature deconfinement phase
transition at
$T_c$. In the low-temperature phase it is thus possible to test
the detailed predictions concerning the Dirac spectrum at small eigenvalues
\cite{VZ,ADMN}. In contrast to the situation in four-dimensional
QCD, the prediction is here that the {\em quenched} microscopic spectral
density should be completely flat.\footnote{This assumes that the quenched
model with staggered quarks can be understood as the $N_f \to 0$ limit of
the theory with an even number of quarks.} To test this, we have performed a
series of measurements of the lowest-lying Dirac operator eigenvalues, using 
the Ritz functional algorithm of ref. \cite{Bunk}.
The density of low-lying eigenvalues at $\beta=12.0$ is shown in
Figure \ref{histogr} for lattice sizes $8^3$ to $14^3$. Except for the
largest lattice we computed the 10 lowest eigenvalues,
on $14^3$ the 16 lowest ones.
There is a clear trend towards a flat 
spectral distribution with increasing lattice volume. The dip very near
$\lambda \sim 0$ is due to finite-size effects, and indeed this dip becomes
more and more narrow as the volume is increased. While a small peak appears
close to $\lambda=0$, its significance is doubtful since it seems to be
smallest on the largest lattice volume. In addition, it is not well
correlated with the distribution of the smallest eigenvalue, as one would
have expected from any genuine oscillatory behavior. From the flat plateaux
we can extract an average value of the spectral density near the origin,
{\it i.e.} $\rho(0)$. We find a slight volume dependence in the extracted
values of $\rho(0)$. Extrapolating as $1/V$ to the infinite-volume limit
we obtain $\rho(0) = 0.0315(19)$. Through a 3-dimensional
analogue of the Banks-Casher formula this provides us with an
independent measurement of the condensate:
$\langle\bar{\psi}\psi\rangle = \pi\rho(0)$. Inserting the above
value for $\rho(0)$, this gives $\langle\bar{\psi}\psi\rangle = 0.099(6)$,
which is the value shown in Figure \ref{pbpvsm}. It is
consistent with our direct measurements. We have thus provided 
yet more independent confirmation of the formation of a condensate in the 
confined phase, 
while simultaneously testing the predictions about
the microscopic spectral density of the Dirac operator \cite{VZ,ADMN}.

\vspace{1cm}
\noindent
{\sc Acknowledgement:} P.H.D. and U.M.H. acknowledge the support of NATO
Science Collaborative Research Grant No. CRG 971487. The work of P.H.D has 
also been partially supported by EU TMR grant no. ERBFMRXCT97-0122, and the 
work of U.M.H. has been supported in part by DOE contracts DE-FG05-85ER250000
and DE-FG05-96ER40979. A.K. acknowledges 
the funding by the Portuguese Funda\c c\~ao para a Ci\^encia e a Tecnologia, 
grants CERN/S/FAE/1111/96 and CERN/P/FAE/1177/97.

\end{document}